\documentclass[12pt,a4paper]{article}

\usepackage{geometry}
\usepackage{authblk}
\usepackage{blindtext}
\usepackage[utf8]{inputenc} % Required for inputting international characters
\usepackage{hyperref}
\usepackage{mathptmx} % Output font encoding for international characters
\usepackage{graphicx}
\usepackage{simpler-wick}
\usepackage{amsmath}
\usepackage{amssymb}
\usepackage{physics}
\usepackage{slashed}
\usepackage{tabu}
\usepackage{array}
\usepackage{float}
\usepackage{wrapfig}
\usepackage{lmodern}
\usepackage[compat=1.0.0]{tikz-feynman}
\usepackage{subcaption}
\newcommand{\keyword}[1]{\textbf{#1}}
\usepackage[autostyle=true]{csquotes} % Required to generate language-dependent quotes in the bibliography

\begin{document}
	
	\title{\textbf{Charged Lepton Flavor Violation in SU(5)  SUSY GUT and MSSM + Type-I Seesaw Model}\vspace*{-3mm}}% Force line breaks with \\
	%\vspace*{-3mm}
	\author[1]{Devidutta Gahan\thanks{ddutta.gahan@gmail.com}}
	%\vspace*{-3mm}
	\author[1]{\vspace*{-3mm} Ng. K. Francis\thanks{francis@tezu.ernet.in}}
	%\vspace*{-3mm}
	\affil[1]{\vspace*{-5mm} \textit{Department of Physics, Tezpur University, Assam, India}}
	\date{}
	\maketitle

\begin{minipage}{0.97\textwidth}
\begin{center}
	\textbf{\fontsize{15}{1} \selectfont Abstract}\\
\end{center}
		Till today lepton flavor violation has not been observed in processes involving charged leptons. Hence, a search for it is under hot pursuit both in theories and experiments. In our current work, we investigate the rates of rare decay processes such as \(\tau \rightarrow \mu \gamma\) in SU(5) SUSY GUT and found that it satisfies the current bound and is one order below the projected sensitivity. This gives a corroborative argument for the influence of the large top-Yukawa coupling at the GUT scale (\(\lambda_{tG}\)) on flavor violating decay rates of leptons which are investigable at low energy electroweak scale \(M_Z\). Secondly, we discuss the decay rates of \(\mu \rightarrow e \gamma\) \& \(\tau \rightarrow \mu \gamma\) in MSSM with added right handed neutrino superfields. From this, we set bounds on \(\tan \beta\) and further, we investigate the mass of \(\tilde{\chi}^0 _1\), the LSP, using the rates of LFV decays. In the calculations, the latest updated data from LHC, neutrino oscillation experiments and constraints on branching ratios from the MEG experiment have been used.
		\begin{description}
			\item[\keyword{Keywords:}] Branching ratios, Supersymmetry, Lepton Flavor Violation(LFV), GUT scale
		\end{description}
\end{minipage}

\section{Introduction}
\hspace*{5mm} Flavor mixing is a well observed phenomenon in the quark sector of the \textit{standard model (SM)}. So also is observed in the neutral lepton sector in form of \textit{neutrino oscillations}. But flavor conservation of charged leptons (LFC) occurs as an accidental symmetry in the SM. No natural theoretical argument exists in favour of LFC. It can be shown that with one loop diagrams (Fig.1) involving a trilinear gauge coupling of \(WW\gamma\) and an oscillating neutrino, the processes are possible in SM. Branching ratio (B.R) for the process of our interest is given as 
\begin{equation}
B.R(\tau \rightarrow \mu\gamma) \backsimeq \frac{\Gamma(\tau \rightarrow \mu\gamma)}{\sum \Gamma_i (\text{all possible decay modes})} \propto \left|\frac{m^2 _{ij}}{M^2 _{W}}\right|^2
\end{equation}
As we can see, the rate is suppressed due to a negligibly small value of the mass squared difference of neutrinos (\(m^2 _{ij} \equiv \Delta m^2 _{ij}\)) on the scale of W-bososn mass. With the current bound on the \(\Delta m^2 _{ij}\) i.e, \(\Delta m^2 _{21} = (7.55 \pm 0.18) \times 10^{-5} eV^2\), \(\left|\Delta m^2 _{31}\right| = (2.50 \pm 0.03) \times 10^{-3} eV^2\) and \(\Delta m^2 _{32} = (2.444 \pm 0.034) \times 10^{-3} eV^2\) from \cite{Ref1,Ref2}, the \(B.R(\tau \rightarrow \mu\gamma) \sim 10^{-54}\), way below the current \& future experimental sensitivity.
	\begin{figure}[t]
		\centering
		\includegraphics[width=0.81\linewidth, height=5cm]{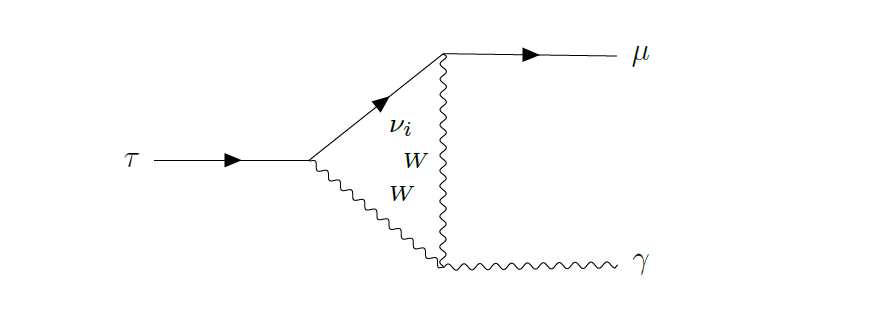}
	\caption{Possible Feynman diagram for decay of \(\tau \rightarrow \mu \gamma\) with all SM couplings}
	\end{figure}

\par But with recent developments in the supersymmetric (SUSY) theories, and especially SUSY Grand Unified Theories (GUTs) \& variations of Minimal Supersymmetric SM (MSSM), interest has grown towards lepton flavor violating (LFV) decays like \(\mu \rightarrow e\gamma\), \(\tau \rightarrow \mu\gamma\), \(\mu \rightarrow eee\) etc, and processes like neutrinoless \(\mu-e\) conversion in a nuclei. The accounting of this phenomenon is linked with a new source of flavor mixing which are mass matrices of the slepton multiplets. In a SUSY-GUT theory one would expect LFV to be quite natural as the leptons and quarks are in same multiplet. The reason that rates of LFV processes predicted by these theories are in reach of the ongoing and upcoming experiments, they are significant testing ground of the SUSY, GUTs and SUSY GUTs. Few works in the past have indicated that LFV processes can be linked with the properties of heavy sterile neutrinos \cite{Ref3}. Hence, probing one can be an indirect search for the other.

\section{Source of Lepton Flavor Violation in a SU(5) SUSY Theory}
\hspace*{5mm} Basic idea of a SUSY theory is loosing discrimination between bosons and fermions by relating them equivalently by transformations. Specifically, weak scale SUSY requires that all SM fermions have scalar super-partners (whose spins differ by \(\frac{1}{2}\)) called  \enquote*{sfermions}, in two categories \enquote*{sleptons} \& \enquote*{squarks}. Also, the SM bosons have fermionic super-partners. Though this doubles the particle content of SM, but allows a number of new physics phenomenology and trivially provides solutions to a number of unaddressed problems in SM. Without any restrictions, a complete general superpotential predicts a rapid violation of global quantum numbers in SM those are lepton and baryon numbers, rendering the theory unphysical. To avoid such an outcome, additional global symmetry called \enquote*{R-parity} (or matter parity) is imposed (See Ch.3 of \cite{Ref4}). With such an imposition the theory conserves baryon number (B), lepton number (L) but allows violation of lepton family number (\(L_i\)) \cite{Ref5}. The predicted rates of these rare decays are relatively low, which have been investigated by many experiments and resulted in null results. Suppression of these rates can be justified with the assumption that the SUSY breaking is flavor blind. This can be further argued in favor by assumption that SUSY is broken in a sector far above the GUT scale \cite{Ref6}, which leaves us with the boundary condition that the slepton-lepton mass matrices are proportional to unit matrices. \\

\par But expectation of the LFV decays are higher in a GUT. The obvious reason is placement of quarks and leptons in the same multiplets. In a minimal SU(5) theory quarks and leptons are arranged in two special multiplets \(\mathbf{\overline{5}}\) \& \(\mathbf{10}\)-plet representation. The CP conjugates of left handed down-quarks (\(d^c _{Li}\)) and left handed leptons (\(e_L\)) are arranged in \(\mathbf{\overline{5}}\)-plet (\(F\)) representation (where \(i\) is the SU(3) index). And \(\mathbf{10}\)-plet representation (\(T\)) contains the SU(2) quark doublet \(Q\), \(u^c _{Li}\) \& \(e^c _L\). And the (\(\mathbf{\overline{5}} \oplus \mathbf{10}\)) representation comes analogously for three generations of matter. A higher unification leads to fewer multiplets, leaving less options for the such rotations which don't generate flavor mixing. Hence, expectation of a higher LFV decay rates is acceptable in this scenario.\\

\par The assumption that SUSY breaking mechanism is flavor blind makes the mass matrices proportional to unit matrices. So mass matrices of different types of fermions can be diagonalized simultaneously, without inducing flavor mixing. But in a supersymmetric unified theory, it is well known that at least two Yukawa type matrix are necesssary to describe masses of fermions. Let's call the matrix \(\boldsymbol{\lambda^u}\) which gives mass to up-quarks in the \(\mathbf{10}\)-plet and \(\boldsymbol{\lambda^d}\) which gives mass to down-quarks \& leptons in the \(\mathbf{\overline{5}}\)-plet representation. Let's choose to work, for the time being, in a basis where the matrix \(\boldsymbol{\lambda^u}\) is diagonal. Call it u-basis. There is an element '\(\lambda_t\)' in this \(\boldsymbol{\lambda^u}\) matrix whose value is very large, accounting for the very heavy mass of top-quark i.e. \(173.1 \pm 0.9\) GeV. The particles sitting in the same multiplet of top quark, one of them is CP conjugate of the 3rd generation lepton \(e^c _{L3}\), will also interact via \(\lambda_t\). Now, we have lost the freedom of diagonalizing mass matrices of sleptons, as leptons and quarks belong to the same multiplet in a SUSY unified theory and we already have chosen a basis to diagonalize \(\boldsymbol{\lambda^u}\). When this theory is scaled down to low energies, using the RGEs, due to significantly large value of \(\lambda_t\), considerable radiative corrections are induced in slepton mass eigenvalues of each generations differently forcing them to be non-degenerate. And beneath unified mass scale, the effective theory is MSSM. Now, as unification of quarks and leptons is no longer there, we can make a relative rotation of the basis, to choose a suitable mass eigenbasis for the leptons, relative to those of quarks \cite{Ref7}. As the masses of sleptons are now non-degenerate, these rotations will sprout non-zero and non-diagonal entries in mass matrices of leptons. These non-zero entries will contribute to the flavor violating decays of heavy leptons making the processes observable at a low energy (i.e. electroweak) scale \(M_Z\).

\section{Lagrangian and scaling the theory}
\hspace*{5mm} Before proceeding further, a few assumptions are made to keep the discussion and calculations fairly simple and straightforward. The SUSY breaking terms are responsible for flavor violation in the leptonic sector. And without any loss of generality it is stated that, soft SUSY breaking is transmitted to matter via supergravity couplings \cite{Ref8}. Notations followed henceforth is closely from \cite{Ref7} with some minor modifications. 
\par As stated before in a SU(5) theory, matter occurs in a (\(\mathbf{\overline{5}} (\overline{F}) \oplus \mathbf{10} (T)\))-plet representation. It is coupled with two higgs supermultiplets, a 5 \((H)\) and a \(\overline{5}\) \(({\overline{H}})\), to form the Yukawa superpotential,
\begin{equation}
W = \tilde{T}^T \boldsymbol{\lambda^u} \tilde{T} H + \tilde{T}^T \boldsymbol{\lambda^d} \overline{\tilde{F}} \overline{H} 
\end{equation}
A tilde in the above expression is indicative that \enquote*{sfermions} are fitted into the SU(5) multiplets. There are additional terms in the above superpotential, but due to their non-direct and non-significant involvment in the light of our discussion, are ignored. Also, a number of terms are uneffective because of knitting R-parity into the theory. Relevant part of the softly broken SUSY Lagrangian is,
\begin{equation}
\mathcal{L}_{soft} = \tilde{T}^{\dagger}\boldsymbol{m}^2 _T \tilde{T} + \overline{\tilde{F}}^{\dagger}\boldsymbol{m}^2 _F \overline{\tilde{F}} + m^2 _H |H|^2 + m^2 _{\overline{H}} |\overline{H}|^2 + \tilde{T}^{T}\boldsymbol{A^u \lambda^u}\tilde{T} H + \tilde{T}^{T}\boldsymbol{A^d \lambda^d}\overline{\tilde{F}} \tilde{H}
\end{equation}
Another assumption is added at this piont, the above superpotential and the softly broken SUSY Lagrangian is valid for an energy range of \(M_G< E < M_{Pl}\). As SUSY is assumed to be broken at maximum energy of the range that is \(M_{Pl}\), so at this energy scale,
\begin{align*}
\boldsymbol{m}^2 _{\tilde{T}} & = \boldsymbol{m}^2 _{\overline{\tilde{F}}} = m^2 _0 \boldsymbol{1}\\ 
m^2 _H & = m^2 _{\overline{H}} = m^2 _0
\end{align*}
\vspace*{-8mm}
\begin{equation}
\hspace*{-2mm}
\boldsymbol{A^u} = \boldsymbol{A^d} = A_0 \boldsymbol{1}
\end{equation}
We have already chosen the basis where \(\boldsymbol{\lambda^u}\) is diagonalized. By working out the RGEs with only one loop effects, the renormalization of the parameters of soft SUSY broken Lagrangian can be achieved. The scaling down to energy scale \(M_G\) becomes simpler when the effects are taken by the terms \(\lambda_t\). The flavor universality and the mass of the fermions start to diverge from degeneracy. But, the \(\boldsymbol{\lambda^u}\) remains diagonal.
\par Now, the theory has need of scaling down further to be accessible by various experiments. Below \(M_G\) the effective theory is MSSM. When the unification is broken, there will be a clear distinction between the \textit{sleptons} \& \textit{squarks}. As GUT is broken, the dimensionality is reduced, a triplet containing SU(2) doublet and a sfermion singlet can be used to write the Lagrangian. Let's look at only the part of Lagrangian concerning to the slepton masses and quark \& lepton masses. 
\begin{itemize}
	\item The slepton mass terms are,
	\begin{equation}
	-\mathcal{L}^{Slepton} _m = \tilde{L}^{\dagger} \boldsymbol{m}^2 \tilde{L} + \tilde{e}^{\dagger} \boldsymbol{m}^2 _e \tilde{e} _R - \tilde{e}^T _R (\boldsymbol{A}^e + \boldsymbol{1} \mu tan\beta) \boldsymbol{\lambda^e}\tilde{e}_L v_d + h.c
	\end{equation}
	\item The quark and lepton mass terms are from Yukawa couplings,
	\begin{equation}
	\mathcal{L}_Y = Q^T \boldsymbol{\lambda}^u _Z u^c _L v_u + Q^T \boldsymbol{\lambda}^d _Z d^c _L v_d + e^{cT} _L \boldsymbol{\lambda}^e _Z L v_d
	\end{equation}
\end{itemize}
An explicit \(\mu\)-parameter is introduced in the slepton mass terms. As we are still in our chosen basis, \(\boldsymbol{\lambda^u}\) is diagonal. But the matrices \(\boldsymbol{\lambda^d}\) \& \(\boldsymbol{\lambda^e}\), which were identical in GUT scale, have been shifted by the effects of renormalization. Parameters \(v_u\) \& \(v_d\) are the vev of the two higgs in the MSSM. The former responsible for mass of up-quarks and the latter for down-quarks and leptons. And \(tan\beta = v_u / v_d\). From the \(\mathcal{L}^{Slepton} _m\), the mass matrices are
\begin{align*}
\boldsymbol{m}^2 _L & = m^2 _L \boldsymbol{1}\\
\boldsymbol{m}^2 _e & = m^2 _e \boldsymbol{1} - \boldsymbol{I}_G
\end{align*}
\vspace*{-8mm}
\begin{equation}
\boldsymbol{A}^e = A_e \boldsymbol{1} - \frac{1}{3} \boldsymbol{I}^{\prime} _G 
\end{equation}
The parameters \(I_G\) \& \(I^{\prime} _G\) are the LFV parameters. These along with \(m^2 _L\), \(m^2 _e\) \& \(A_e\) can be derived using the RGEs. 
\par Now let's diagonalize \(\boldsymbol{\lambda^d} _Z\) \& \(\boldsymbol{\lambda^e} _Z\). We have
\begin{align}
\boldsymbol{\lambda^d} _Z \times v_d & = \boldsymbol{V}^{*} \boldsymbol{M}^d \boldsymbol{U}\\
\boldsymbol{\lambda^e} _Z \times v_d & = \boldsymbol{V}^{e*} \boldsymbol{M}^e \boldsymbol{U}^{e\dagger}
\end{align}
Where, \(\boldsymbol{M}^d\) \& \(\boldsymbol{M}^e\) are the diagonalized mass matrices for down-quarks and leptons and \(\boldsymbol{U}\) \& \(\boldsymbol{V}\) are the good old CKM-matrices. But \(\boldsymbol{V}^e\) is an extended CKM-matrix. The elements of \(\boldsymbol{V}^e\) are related to \(\boldsymbol{V}\) by the relations \cite{Ref9},
\begin{equation}
\boldsymbol{V}^e _{ij} = y\boldsymbol{V} _{ij} \text{ for } i\neq j; \text{ \& } i\text{ or } j =3;  \boldsymbol{V}^e _{ij} = \boldsymbol{V} _{ij}  \text{ (otherwise)}
\end{equation}
The expression for \(y\) will be stated in the next section. As the extended CKM-matrix is used to diagonalize mass matrix of the charged leptons, we expect non-zero and non-diagonal entries in this mass matrix. This will induce significant rates of LFV processes. Using ideas explained in this section we will calculate the rate for \(B.R(\tau \rightarrow \mu \gamma)\).

\section{Calculating \(\tau \rightarrow \mu \gamma\) in SU(5) SUSY GUT}
We use the following relations which can be used to calculate the \(B.R(\tau \rightarrow \mu\gamma)\) in a supersymmetric SU(5) theory.
\begin{equation}
\frac{B.R(\tau \rightarrow \mu\gamma)}{B.R(\mu \rightarrow e\gamma)}\Bigg|_{SU(5)} = \left|\frac{\mathbf{V^e _{\tau\tau}}}{\mathbf{V^e _{e\tau}}}\right|^2 \times B.R(\tau \rightarrow \mu \nu \overline{\nu})
\end{equation}
where the matrix elements are from the extended-CKM matrix defined above are
\begin{equation}
\mathbf{V^e _{\tau \tau}}= \mathbf{V_{tb}}  \hspace{3mm} \text{\&} \hspace{3mm} \mathbf{V^e _{e\tau}} = y\mathbf{V_{td}} 
\end{equation}
We are going to use the updated values of the CKM-matrix, to calculate the rate. The expression for '\(y\)' is derived from the RGE for MSSM. 
\begin{equation}
y(E) \equiv  exp\Big[-\int_{lnE}^{lnM_G} \frac{\lambda^2 _t (E^{\prime})}{16\pi^2} d(lnE^{\prime})\Big] = [1-\rho(E)]^{1/2b_t}
\end{equation}
where
\begin{equation}
\rho(E) \equiv \frac{\lambda^2 _t (E)}{(\lambda^2 _t)^{max} (E)}
\end{equation}
We adopt to put a subscript 'Z' with the parameters in-order to indicate that they are evaluated at \(E = M_Z\). The dependency expressions are,
\begin{align}
\lambda^2 _t (E) &  = \frac{(\lambda^2)^{max} _t (E)}{1+\frac{(\lambda^2)^{max} _t (E)}{\lambda^2 _{tG}J_u (E)}}\\
J_{\alpha} (E) & = \prod_{i} f^{c^{\alpha} _i / b_i} _i (E)\\
f^{c^{\alpha} _i / b_i} _i (E) & = 1 + b_i g^2 _G t(E)\\
t(E) & = \frac{1}{(4\pi)^2} ln\frac{M^2 _G}{E^2}\\
\text{where } & i=1,2,3 \text{ \& } \alpha =u,d,e.
\end{align}
\begin{table}[t]
	\centering
	\begin{tabular}{||c|| c | c c c| c c c ||} 
		\hline
		\(i,g\) & \(b_i\) & \(c^u _i\) & \(c^d _i\) & \(c^s _i\) & \(b^u _g\) & \(b^d _g\) & \(b^s _g\) \\ [0.7ex] 
		\hline
		 1 & \(\frac{33}{5}\) & \(\frac{13}{15}\) & \(\frac{7}{15}\) & \(\frac{9}{5}\) & 3 & 0 & 0\\
		 2 & 1 & \(3\) & \(3\) & \(3\) & 3 & 0 & 0\\
		 3 & \(-3\) & \(\frac{16}{3}\) & \(\frac{16}{3}\) & \(0\) & 6 & 1 & 0\\
		\hline
	\end{tabular}
	\caption{Values of RGE coefficients in MSSM}
\end{table} 
From the ratio between \((\lambda^2)^{max} _t (E)\) \& \(\lambda^2 _{t}(E)\), we write the expression for \(\rho(E)\) as
\begin{flushleft}
	\begin{align*}
	\rho(E) & = \frac{\lambda^2 _t (E)}{(\lambda^2 _t)^{max} (E)} =  \frac{1}{1+\frac{(\lambda^2)^{max} _t (E)}{\lambda^2 _{tG}J_u (E)}}
	\end{align*}
	\begin{equation}
	\rho_Z = \rho(M_Z) = \frac{1}{1+\frac{(\lambda^2)^{max} _t (M_Z)}{\lambda^2 _{tG}J_u (M_Z)}}
	\end{equation}
\end{flushleft}
We first determine the \(t_Z \equiv t(M_Z)\). Most recent value of \(M_Z = 91.1876 \pm 0.0021\) GeV and \(M_G \sim 2 \times 10^{16}\) GeV \cite{Ref1}. This gives
\[
t_Z = \frac{1}{(4\pi)^2} ln\frac{(2 \times 10^{16})^2}{(91.1876)^2} = 0.41822
\]
The improved value of \(g_G = \sqrt{4\pi \alpha_G}\), from \(\alpha^{-1} _G = 24.3\), is \(g_G = 0.7191\). Values of the RGE coefficients in MSSM from Table 1 give \(f_{1Z} = 2.4273\), \(f_{2Z} = 1.2162\) \& \(f_{3Z} = 0.3512\). Using them we get
\[
J_{uZ} = (f_{1Z})^{c^u _1 / b_1}\cdot(f_{2Z})^{c^u _2 / b_2} \cdot (f_{3Z})^{c^u _3 / b_3} = 12.9849
\]
Calculating the value of \((\lambda^2)^{max} _t (E)\) analytically is not possible. It has to be calculated numerically \cite{Ref10}, which turns \(\lambda^{max} _{tZ} = 1.14\). The last piece that we need to calculate \(\rho_Z\) is \(\lambda_{tG}\). The value is linked to \(\lambda^{max} _{tG}\), we take \(\lambda^2_{tG}=0.9 (\lambda^2)^{max} _{tG}\). The value of \(\lambda^{max} _{tG}\) can be determined from the RGE of the GUT scale with \(E_{max} = M_{Pl} \sim 1.22 \times 10^{19} GeV\) \cite{Ref1}. 
\begin{equation}
(\lambda^2)^{max} _t (E) = \frac{c^u + b_G}{b_t} \frac{g^2 _5 (E)}{1-[\frac{f_5 (E_{max})}{f_5 (E)}]^{1+c^u / b_G}}
\end{equation}
where
\begin{align}
& f_5 (E) = 1+g^2 _G b_G [t(E) - t(M_G)]\\
& t(E) =  \frac{1}{(4\pi)^2} ln\frac{M^2 _{Pl}}{E^2}
\end{align}
We will put a subscript \enquote{\(G\)} to indicate a parameter evaluated at \(E=M_G\). As, \(g_G \equiv g_5 (M_G)\). The parameter \(b_G\) is the \(\beta-\)function coefficient, which is chosen to be \(b_G = -3\). And \(b^u _3 \equiv b_t\) after evaluating the various parameters, we get \(t_G = 0.04046\), \(f_{5G} = 1\), \(f_5 (E_{max} = f_5 (M_{Pl}) = 1.0629\). The RGE coefficients in SU(5) (\(c^u = \frac{96}{5}, b_t = 9, b_G = -3\)). This gives
\begin{align*}
(\lambda^2)^{max} _{tG} & =(\lambda^2)^{max} _t (M_G) = \frac{\frac{96}{3} + (-3)}{9} \cdot \frac{0.7191^2}{1- [1.0629]^{1+\frac{96}{-15}}}\\
\Rightarrow & (\lambda^2)^{max} _{tG} = 3.3165\\
\Rightarrow & \lambda^{max} _{tG} = 1.8211
\end{align*}
We have taken \(\lambda_{tG} = \sqrt{0.9} \lambda^{max} _{tG}\), which gives the value
\[
\lambda_{tG} = 1.7276
\]
Using this, the value for \(\rho_Z = 0.9675\), correspondingly \(y =  0.7516\).
\par Now we come to the prediction of branching ratios. We take most recent \& updated values from \cite{Ref1} to calculate the results. \(B.R(\tau \rightarrow \mu \nu \overline{\nu}) = 0.1733^{+0.0005} _{-0.0005}\). The current upper limit of the \(B.R.(\mu \rightarrow e\gamma)<4.2\times 10^{-13}\) \cite{Ref14}. Next we take different values of CKM-matrix elements.
\begin{itemize}
	\item The value of \(V_{td} = (8.1\pm 0.5)\times 10^{-3}\). At a confidence level 95\%, \(0.90< V_{tb} < 0.99\) was obtained from the data collected at the end of Run II of the Tevatron, jointly from D\(\slashed{O}\) and CDF \cite{Ref11}. The same quantity was reported to be \(0.975< V_{tb}\) from the 8 TeV measurments of CMS \cite{Ref12}. Refining the range \(0.975<V_{tb}<0.99\), and averaging the same we fix the value at \(V_{tb} = 0.9825\) for our calculation.
	\begin{subequations}
		\begin{equation}
			B.R(\tau \rightarrow \mu \gamma)|_{SU(5)} < 1.67 \times 10^{-9} 
		\end{equation}
	\item Keeping the same value of \(V_{td}\), we use the constraint on the \(V_{tb}\) obtained from precision electroweak data, where the top quark enters a loop. The sensitivity is best in \(\Gamma(Z\rightarrow b\overline{b})\) reported the value \(V_{td} = 0.77^{+0.18} _{-0.24}\) \cite{Ref13}.
	\begin{equation}
	B.R(\tau \rightarrow \mu \gamma)|_{SU(5)} < 6.2535 \times 10^{-10}
	\end{equation}
	\item Using the global fit of all the available data, and imposing the constraints in the SM, one can determine the CKM matrix elements most accurately, which gives \(V_{td} = 0.00896^{+0.00024}  _{-0.00023}\) \& \(V_{tb} = 0.999105 \pm 0.000032\) \cite{Ref1}.
	\begin{equation}
	B.R(\tau \rightarrow \mu \gamma)|_{SU(5)} < 1.515\times 10^{-9}
	\end{equation}
	\end{subequations}
\end{itemize}

\section{LFV in MSSM with Right Handed Neutrinos}
\hspace*{5mm} In this section we will move our attention to the Seesaw Type-I \cite{Ref15,Ref16} combined with MSSM \cite{Ref17}. The idea of LFV in a SUSY GUT happens to be quite intuitive because leptons and quarks are contained in the same multiplets. But, as the SM fermions turn to be scalars in SUSY models, same idea of LFV in them becomes less intuitive. Seesaw models, theoretically, provide a natural framework for explaining mass and oscillations of left handed active neutrinos by introducing right-handed SU(2) singlet neutrino field (sterile neutrinos) of heavier mass into the theory. Following the same prescription, the field content for Type-I Seesaw combined with MSSM will include an addition of a \keyword{Right Handed Neutrino (RHN)} superfield for each generation in the MSSM.
\begin{equation*}
\tilde{L}: \begin{pmatrix}
1,2,-\frac{1}{2}
\end{pmatrix}; \hspace{3mm}
\tilde{e}^{c} : \begin{pmatrix}
1, 1, +1
\end{pmatrix}; \hspace{3mm}
\tilde{\nu}^{c} : \begin{pmatrix}
1, 1, 0
\end{pmatrix}
\end{equation*}
\vspace*{-5mm}
\begin{equation}
\tilde{Q} : \begin{pmatrix}
1,2,+\frac{1}{6}
\end{pmatrix}; \hspace{3mm}
\tilde{u}^{c} : \begin{pmatrix}
\overline{3}, 1, -\frac{2}{3}
\end{pmatrix}; \hspace{3mm}
\tilde{d}^{c} : \begin{pmatrix}
\overline{3}, 1, \frac{1}{3}
\end{pmatrix}
\end{equation}
\begin{equation*}
\tilde{H}_{u}: \begin{pmatrix}
1,2, \frac{1}{2}
\end{pmatrix}; \hspace{3mm}
\tilde{H}_{d}: \begin{pmatrix}
1,2,-\frac{1}{2}
\end{pmatrix} \hspace{3mm}
\end{equation*}
The \(\tilde{\nu}^{c}\) is the newly added RHN superfield in the theory which happens to be a \(SU(2)\) singlet. The superpotential of such a field content is given as 
\begin{equation}
\mathcal{W} = \mathbf{Y^{d} _{ij}} \tilde{d}^c _i \tilde{Q}_j \tilde{H}_d + \mathbf{Y^{u} _{ij}} \tilde{u}^c _i \tilde{Q}_j \tilde{H}_u + \mathbf{Y^{e} _{ij}} \tilde{e}^c _i \tilde{L}_j \tilde{H}_d + \mu \tilde{H}_u \tilde{H}_d + \mathbf{Y^{\nu} _{ij}} \tilde{\nu}^c _i \tilde{L}_j \tilde{H}_u -\frac{1}{2} M_{R_i}\tilde{\nu}^c _i \tilde{\nu}^c _i
\end{equation} 
Here, except for the ususal \(\mathcal{W}_{MSSM}\) extra two terms get added due to the Type-I Seesaw mechanism. The \(\mathbf{Y^{\nu} _{ij}} \tilde{\nu}^c _i \tilde{L}_j \tilde{H}_u\) is the Dirac type Yukawa coupling which is now possible because of the RHN coupled with the left handed lepton field via the Higgs field used to generate mass of the up-type quarks in our theory. Second, \(\frac{1}{2} M_{R_i} \tilde{\nu}^c _i \tilde{\nu}^c _i\), is the Majorana Mass term for the RHN field added. This superpotential is valid at a scale \(q^2 > M^2 _{R}\) which is often referred to as the \keyword{Seesaw Scale}. Below this scale, the RHNs decouple at their respective mass scales giving rise to a new superpotential containing the \(\mathcal{W}_{MSSM}\) plus an higher-dimensional operator.
\begin{equation}
\mathcal{W}_{decoupled} = \mathcal{W}_{MSSM} + \frac{1}{2}(\mathbf{Y^{\nu}} L H_u)^{T} M^{-1} _N \frac{1}{2}(\mathbf{Y^{\nu}} L H_u)
\end{equation}
The mass states \(M_N\) are sterile states (in a SUSY context) which are generated after the soft-SUSY breaking takes place and the Lagrangian takes the form
\begin{equation}
-\mathcal{L}_{soft, Seesaw} =  -\mathcal{L}_{soft} + (m^2 _{\tilde{\nu}})^i _j \tilde{\nu}^{*} _{Ri} \tilde{\nu}^{j} _{R} + (\frac{1}{2} B^{ij} _{M_{\nu}} \tilde{\nu}^{*} _{Ri} \tilde{\nu}^{*} _{Rj} + A^{ij} _{\nu} \tilde{L} _i \tilde{\nu}^{*} _{Rj})
\end{equation}
It becomes very crucial that we point out a clear distinction, these sterile states (mass \(M_N\)) are not the same sterile neutrinos whose searches are ongoing at various experiments \cite{Ref18} these are SUSY states and have very high mass as we shall see in the next sections. After their decoupling, we have the same particle content as that of MSSM at low energy. This digression, naturally explains the neutrino oscillations and light neutrino masses at the EW scale, which is given by an effective Majorana mass matrix
\begin{equation}
m_{eff} = -\frac{1}{2} \langle H^0 _u\rangle^2 Y^{\nu} M^{-1} _N (Y^{\nu})^T
\end{equation}

\par In the Eq.(26) we can observe that the fields \(L_j\) \& \(\nu^{c} _i\) can be rotated in such a way that the lepton Yukawa couplings \(Y^{e} _{ij}\) and RHN Majorana mass matrix \(M_{Ri}\) can be diagonal. But in such a basis the neutrino Yukawa couplings \(Y^{e} _{ij}\) are non-diagonal in general. And this is what gives rise to the LFV effects which can be observed at low energy \cite{Ref20,Ref21,Ref22}. As stated earlier in Section 1.1, the LFC is not an implication or restriction imposed by the gauge symmetries rather occur as accidental symmetry in SM. Hence, the slepton mass matrices can induce LFV without being inconsistant with th gauge symmetry. Infact, they can be very good observable at the EW scale, much below the RHN scale \(M_R\). The soft slepton mass matrices acquire contributions from  RGEs, running from \(M_{GUT}\) to \(M_R\), into their off diagonal elements which are stated below in their leading order log-approximation \cite{Ref23}, as
\begin{subequations}
\begin{equation}
(m^2 _{\tilde{L}})_{ij} \sim \frac{1}{16 \pi^2} (6m^2 _0 + 2A^2 _0) (Y^{\nu \dagger} Y^{\nu})_{ij} \log(\frac{M_{GUT}}{M_{R}}) 
\end{equation}
\begin{equation}
(m^2 _{\tilde{e}})_{ij} \sim 0
\end{equation}
\begin{equation}
(A_l)_{ij} \sim \frac{3}{8\pi^2} A_0 (Y_L)_{ij} ((Y^{\nu \dagger} Y^{\nu})_{ij}) \log (\frac{M_{GUT}}{M_{R}})
\end{equation}
\end{subequations}

\section{Calculation and Results from the MSSM+Type-I Seesaw}
\hspace*{5mm} Using the framework of \texttt{SuSeFLAV 1.2} \cite{Ref24} the rates of LFV processes particularly \(\mu \rightarrow e \gamma\) \& \(\tau \rightarrow \mu \gamma\) are calculated. The framework is able to run a number of variants with a wide variety of options for the users like the specific mechanism of SUSY breaking, whether or not to include the RHN, their masses, the mass scale of SUSY breaking, mixing type of RHN, etc,. The framework of \texttt{SuSeFLAV 1.2} has two modes of running one is with single point input and single point output and the second is parameter space scanning mode. Our analysis is based on the Minimal Supergravity (\textbf{mSUGRA}) \cite{Ref25} SUSY breaking mechanism, in which all the phenomenologies can be characterized by four parameters \(m_0, A_0, m_{1/2}, \tan \beta\) and sign of \(\mu\). The primary reason behind such a choice is that it exibhits minimal flavor violation (MFV) \cite{Ref26,Ref27} because of the flavor blind soft-SUSY terms. Whatever the flavor violation couplings are generated, only because of the RGEs running between the EW scale and the \(M_{GUT}\) scale, which is chosen to be the scale of SUSY breaking in our work. The description provided by this picture of MFV fits better with physical observations made by the experiments yet and the predictions are investigable by the upcoming experiments. We used the framework in scanning mode to analyze the predictions of this theory about the rates of LFV processes. The parameter space is of 4 continuous variables: \(m_0, A_0, m_{1/2}\) \& \(\tan \beta\). Our parameter space is inspired from \cite{Ref28}, for scanning is listed in table below.
\begin{table}[t]
	\centering
	\fontsize{12pt}{14}\selectfont
	\begin{tabular}{ | m {5cm} || m {5cm}|}
		%\hline
		%\multicolumn{}{|c|}{Country List} \\
		\hline
		Parameter Name & Run-I \\
		\hline
		\(m_0\) & \(100GeV \leq m_0 \leq 10TeV\)\\
		\(M_{1/2}\)&  \(50GeV \leq M_{1/2} \leq 10TeV\) \\
		\(A_0\) & \(-10TeV \leq A_{0} \leq 10TeV\)\\
		\(tan \beta\) & \(2 \leq \tan \beta \leq 65\) \\
		\(\mu\) & \(\mu>0\) \& \(\mu<0\)\\
		\hline
	\end{tabular}
	\caption{The parameter space for Run-I \& II of our analysis}
\end{table} 
A list of SM input parameters at EW scale with their values \cite{Ref1} are taken 
\begin{itemize}
	\item Mass of the top-quark: \(m_t = 173.2 \) GeV
	\vspace*{-1.5mm}
	\item Mass of bottom-quark: \(m_b = 4.18 \) GeV
	\vspace*{-1.5mm}
	\item Mass of tau lepton: \(m_{\tau} = 1.776 \) GeV
	\vspace*{-1.5mm}
	\item Mass of Z-boson: \(M_Z = 91.1876 \) GeV
	\vspace*{-1.5mm}
	\item Inverse of QED coupling costant at \(M_z\): \(\alpha^{-1} _{em} = 137.036\)
	\vspace*{-1.5mm}
	\item The strong coupling contant at \(M_Z\): \(\alpha_s = 0.1181\)
	\vspace*{-1.5mm}
\end{itemize}
The framework specific options are specified in our analysis as follows
\begin{itemize} 
	\item Mixing Case: PMNS 
	\vspace*{-1.5mm}
	\item No. of loop(s) in RGEs': 2-loops
	\vspace*{-1.5mm}
	\item Quark Mixing: Yes
	\vspace*{-1.5mm}
	\item Tolerance of the spectrum: \(10^{-4}\)
	\vspace*{-1.5mm}
	\item Masses of RHN (in GeV): \{\(M_{R_3} ,M_{R_2} ,M_{R_1}\) \} = \{\(10^{14}, 10^9, 10^6\)\} 
	\vspace*{-1.5mm}
\end{itemize}
\par Here we move to our primary objective to investigate the rates of LFV in our model. But the aim that we set for us in this framework analysis is to constraint some parameters of SUSY. First we shall constraint the \(\tan \beta\) by plotting the rates of \(\mu \rightarrow e \gamma\) \& \(\tau \rightarrow \mu \gamma\) v/s \(\tan \beta\). This plot is done by taking the whole parameter space and scanned with 10000 points for values of \(\tan \beta =[2,30] \). This will be referred to as Run-I.
\begin{figure}[!b]
	\centering
	\includegraphics[width=0.92\linewidth, height=7.3cm]{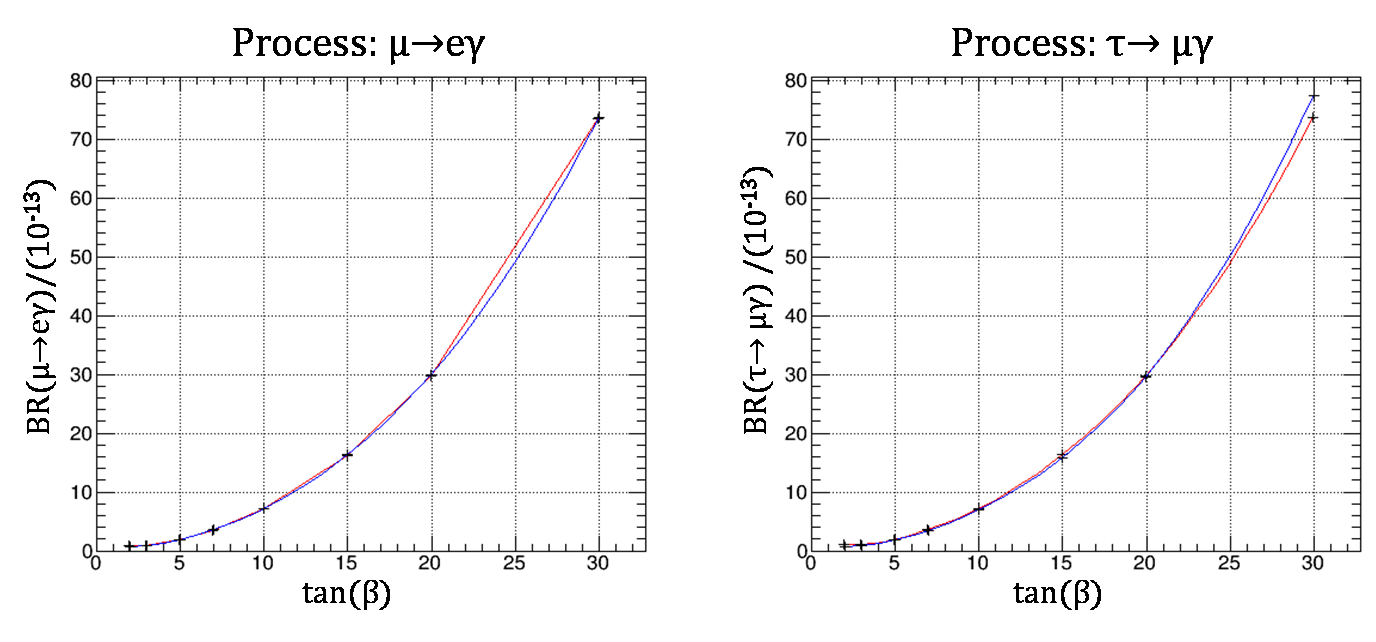}
	\caption{Run-I outputs after the parameter space scanning with 10000 points for each value of \(tan \beta\) and in both the figures line in blue is for \(\mu > 0\) and red is for \(\mu < 0\)}
\end{figure}
\par It is clearly seen from the Fig.2 that in the given parameter space the rates of LFV processes are higher for high values of \(\tan \beta\). Though the initial range of \(\tan \beta\) from Table 2, was wide but in our analysis it can be observed that the rates produced with \(\tan \beta > 10\) are unacceptable because those values have been previously investigated and negated by the experiments. Also for values \(2 \leq \tan \beta \leq 10\) the rates of \(\mu \rightarrow e \gamma\) are near and below the current bound and projected sensitivity \cite{Ref14}. But if we turn our attention to the process \(\tau \rightarrow \mu \gamma\) for the whole parameter space, the rates are well below the current bound and projected sensitivity \cite{Ref29,Ref30}. Here we can comment that the process \(\mu \rightarrow e \gamma\) is a better probe for constraining the theory and parameter space. So for our Run-II (whose purpose will be explained shortly), we shall consider only one process, i.e., \(\mu \rightarrow e \gamma\) within the reduced range \(2 \leq \tan \beta < 10\). This we will section to \(\tan \beta = 2,3,5,7\) and our Run-II scan will consist of the rest of parameter space formed by the remaining 3 parameters.
\par The Run-II has two primary aims, one is to reduce or chop-off the parameter space with an upper and lower bounds and second is to put a limit to the mass of Lightest Supersymmetric Particle (LSP) (which is \(\tilde{\chi}^0 _1\)) using the rates for LFV processes. The Run-II was done in a number of phases, based on number of points in the run. The Phase-1 which scanned the complete parameter space, was done with 100 points and 10 best small ranged parameter spaces were choosen. The criteria adopted in this run to chop-off the parameter space is, we favor the \(\mathcal{O}(10^{-15}) < BR < \mathcal{O}(10^{-13})\) and select points and neighbourhoods accordingly. For Phase-2, we scanned each of these narrowed region of parameter space with 1000 points each and applied the same rejection technique to obtain 3 more favored region. In Phase-3, we scanned these 3 regions again with 1000 point, and observed which region contained most number of rates in the favored region of \(BR\). From here we chose the one region of parameter space with the highest number of favored points which is listed in the Table 3.
\begin{table}[t]
	\centering
	\fontsize{12pt}{15}\selectfont
	\begin{tabular}{ | m {5cm} || m {5cm}|}
		%\hline
		%\multicolumn{}{|c|}{Country List} \\
		\hline
		Parameter Name & After Phase-3, Run-II \\
		\hline
		\(m_0\) & \(3TeV \leq m_0 \leq 5TeV\)\\
		\(M_{1/2}\)&  \(2TeV \leq M_{1/2} \leq 7TeV\) \\
		\(A_0\) & \(2TeV \leq A_{0} \leq 4TeV\)\\
		\hline
	\end{tabular}
	\caption{Our final chopped-off parameter space}
\end{table}

With this parameter space, the final run was done, and then with random selection of points in the output and varying \(\tan \beta\), as mentioned in the beginning of the Run-II, isoplots were done for \(BR(\mu \rightarrow e \gamma)\) v/s \(M_{\tilde{\chi}^0 _1}\).

\begin{figure}[t]
	\centering
	\includegraphics[width=14cm, height=9.3cm]{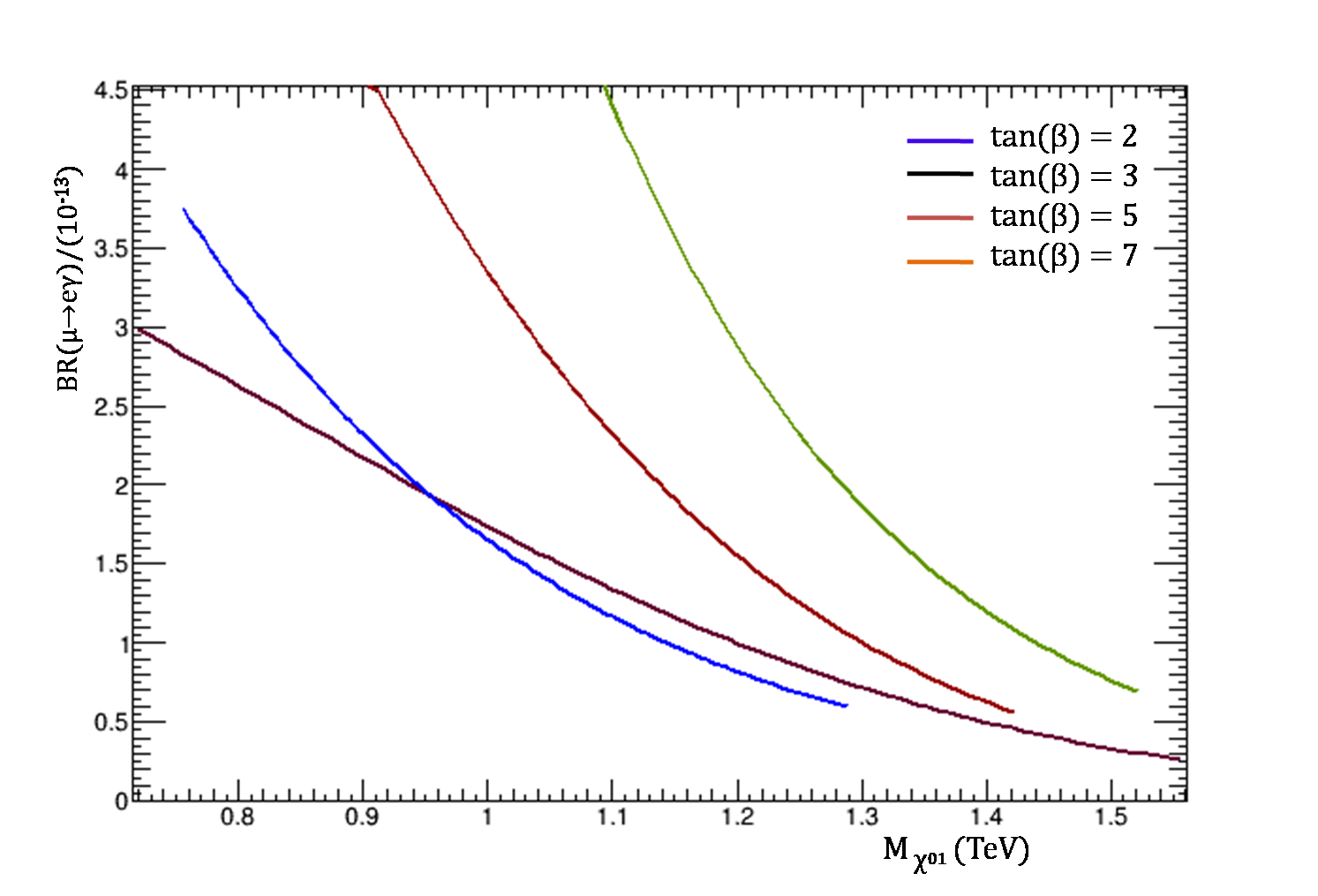}
	\caption{Run-II outputs, isoplot of \(BR(\mu \rightarrow e \gamma)\) v/s \(M_{\tilde{\chi}^0 _1}\)}
\end{figure}

\hspace*{5mm} As already stated earlier, \(\tilde{\chi}^0 _1\) is the LSP. If R-parity is conserved in MSSM, then the LSP is stable and often assumed to be Weakly Interacting Massive Particle (WIMP) \cite{Ref31,Ref32}. So to say, is a great candidate for particle DM. This is why we explored the mass of LSP. The current upper limit on the BR(\(\mu \rightarrow e \gamma\)) stands at \(\sim \mathcal{O}(10^{-13})\) and if we observe our graph in combination with previously used exclusion technique of \(\mathcal{O}(10^{-15}) < BR < \mathcal{O}(10^{-13})\) we can do a direct inference of \(0.75 \text{ TeV} < M_{\tilde{\chi}^0 _1}< 1.5 \text{ TeV}\) for \(\tan \beta = 2,3\). The limit we derived is in good agreement with various searches for the LSP. The Detector with Lepton, Photon and Hadron Identification (DELPHI) Collaboration \cite{Ref33}, which looked at the collision data of \(e^{-}e^{+}\) at center of mass energy of 208-GeV, interpreted it in the context of MSSM with R-parity conservation, searching for Charginos, Neutralinos and sfermions. The analysis set a lower limit of 46-GeV for a stable neutralino mass. But, with new data from LHC and constraining model like Constrained MSSM (CMSSM) \cite{Ref34} a stronger lower bound is set well above 200-GeV for the mass of LSP \cite{Ref35}. The Global and Modular Beyond-the-Standard-Model Inference Tool (GAMBIT) collaboration, post global fit, gave the best fit mass of LSP neutralino to be \(m_{LSP} \simeq 600\)GeV in CMSSM, \(m_{LSP} \simeq 850\)GeV in Non-Universal Higgs Model 1 (NUHM1) and \(m_{LSP} \simeq 930\)GeV in NUHM2. Where, \(LSP = \tilde{\chi}^0 _1\) and the acronyms indicate different manifestations of MSSM. Direct searches of Neutralino and Chargino at LHC with \(\sqrt{s} = 8\)TeV (2014) via Higgs, W and Z bonons in final state with missing transverse momentum (\(\slashed{E}^{miss} _T\)), yield a lower limit of 380 GeV on the mass of \(\tilde{\chi}^0 _1\) \cite{Ref36} from CMS Collaboration. This search was based on Gauge Mediated SUSY Breaking (GMSB) \cite{Ref37} with R-parity conservation. In 2018, another result was published by CMS collaboration \cite{Ref38} with data of \(pp\) collisions at \(\sqrt{s} = 13\) TeV and delayed photons. With the same underlying model assumptions as before, this resulted in a higher lower limit of 525 GeV on the mass of \(\tilde{\chi}^0 _1\). Other analysis like the combined searches by CMS Collaboration provides a lower limit as high as 750 GeV \cite{Ref39} and searches with a lepton, a photon and \(\slashed{E}^{miss} _T\) provides a limit of 930 GeV on the LSP mass \cite{Ref40}. 

\section{Conclusions}
\hspace*{5mm} We have satisfactorily justfied that, a search for Flavor Violating decays of heavier charged leptons are a reliable testing ground for various SUSY theories, whichever we want to establish as a versatile BSM theory. The results obtained in Section 4, after the calculations, are found to have a good agreement with the current sensitivity of \(B.R(\tau \rightarrow \mu \gamma) < 4.4 \times 10^{-8}\) \cite{Ref29} \& projected sensitivity of \(B.R(\tau \rightarrow \mu \gamma) = 4.4 \times 10^{-9}\) \cite{Ref30}. These are achieved by exploring the SU(5) SUSY GUT. The next Section 6, the rates of LFV decays in the MSSM + Type-I Seesaw were explored, with mSUGRA. We saw that in this particular form of SUSY the rates of \(\tau \rightarrow \mu \gamma\) are not a good probe for the theory. And the calculated rates of \(\mu \rightarrow e \gamma\) are in a region agreement with the current bound and investigable by the upcoming experiments (like MEG-II, SHiP, Belle-III, etc.,). Further we went on to investigate \(\tan \beta\) which is a very important parameter in the MSSM to explore a great amount of phenomenologies. If charged LFV decays are assumed then a lower values of \(\tan \beta\) should be preferred lying in the range of 2 to 7. Next, we did an analysis of Neutralino LSP mass, compatible with the current limit of LFV BR. This yields a mass range for \(\tilde{\chi}^0 _1\) to be \(0.75 < m_{\tilde{\chi}^0 _1} < 1.5\) TeV. This is well in agreement with the various searches and global fits done with \(e^{-}e^{+}\) and \(pp\) collisions data. If we again combine the various exclusion limits presented in the \textbf{Table 110.2} of \cite{Ref1}, for squarks and gluinos with R-parity conservation, we can trim the mass scale at 1.1 TeV for best implications. All the limits presented here are pointing towards an R-parity conserving scenario. But, in contrast to the initial assumption of a stable LSP neutralino, majority of the collider data besides our theoretical calculation are pointing towards an unstable Neutralino making it unsuitable for a Dark Matter candidate. The LFV phenomenology can also be combined with the Neutrino Oscillation data to get some new limits on various BSM parameters. So, finally by combining mSUGRA, CMSSM, GMSB \& NUHM 1,2 we can write the unstable LSP neutralino mass to be in the limit of \(0.75 < m_{\tilde{\chi}^0 _1} < 1.1\) TeV, with the assumption of R-parity conserving in all the models. Though we are still waiting for a direct observation of such a decay, but the acquired bounds are still able to help us redefine limits on the values of different MSSM parameters.

\section*{Acknowledgements}
\hspace*{5mm} We thank Sudhir Vempati (CHEP, IISc, Bangalore) and Debtosh Chowdhury (IIT, Kanpur) for help on \texttt{SuSeFLAV 1.2} \cite{Ref24}. Their help in getting acquainted with the framework and running codes quickly was crucial for the progress of this work. This work was supported by DST-SERB, Govt. of India vide research grant no. EMR/2015/001683, for project entitled \enquote{Neutrino Mass Ordering, Leptonic CP-violation and matter-antimatter asymmetry.}


\begin{thebibliography}{9}
	\bibitem{Ref1} 
	M. Tanabashi et al. (Particle Data Group), \textbf{Phys. Rev. D 98 (2018) 030001 and 2019 update.}
	
	\bibitem{Ref2} 
	P. F. de Salas et al., \textbf{Phys. Lett. B782 (2018) 633 }, [arXiv:1708.01186].
	
	\bibitem{Ref3} 
	L. Lello, D. Boyanovsky, \textbf{Nuclear Physics B 880 (2014) 109.}
	
	\bibitem{Ref4}
	P. Ramadevi et. al, \textit{Surveys in Theoretical High Energy Physics 1}, Hindustan Book Agency(India) 2013.
	
	\bibitem{Ref5}
	R. Barbieri and L.J. Hall, \textbf{Phys. Lett. B 338 (1994) 212.}
	
	\bibitem{Ref6}
	R. Barbieri, S. Fert-ara and C. Savoy, \textbf{Phys. Lett. B I10 (1982) 343.}
	
	\bibitem{Ref7}
	R. Barbieri, L.J. Hall, and A. Strumia, \textbf{Nucl. Phys. B 445 (1995) 219.}
	
	\bibitem{Ref8}
	L.J. Hall, J. Lykken and S. Weinberg, \textbf{Phys. Rev. D27 (1983) 2359.}
	
	\bibitem{Ref9}
	M. Olechowski and S. Pokorski, \textbf{Phys. Lett. B 257 (1991) 388.}
	
	\bibitem{Ref10}
	 A. Bouquet, J. Kaplan and C.A. Savoy, \textbf{Nucl. Phys. B 262 (1985) 299.}
	
	\bibitem{Ref11}
	V. M. Abazov et al. The D\(0\) Collaboration , \textbf{Phys. Rev. Lett. 107 (2011) 121802,} [arXiv:hep-ex/1106.5436].
	
	\bibitem{Ref12}
	 V. Khachatryan et al. The CMS Collaboration, \textbf{Phys. Lett. B 736 (2014) 33,} [arXiv:hep-ex/1404.2292].
	 
	\bibitem{Ref13}
	J. Swain and L. Taylor, \textbf{Phys. Rev. D 58 (1998) 093006,} [arXiv:hep-ph/9712420].
	
	\bibitem{Ref14}
	S. Kobayashi, Physical Society of Japan, 2019 Fall Meeting, Yamagata University.\\
	\url{https://meg.web.psi.ch/docs/talks/JPS/2019a/kobayashi_jps2019a.pdf}
	
	\bibitem{Ref15}
	 P. Minkowski, \textbf{Phys. Lett. B 67 (1977) 421.}
	
	\bibitem{Ref16}
	 R. Mohapatra and G. Senjanovic, \textbf{Phys. Rev. Lett. 44 (1980) 912.}
	
	\bibitem{Ref17}
	S. P. Martin, \enquote{Supersymmetry primer}, \textbf{[arXiv:hep-ph/9709356]}.
	
	\bibitem{Ref18} S. Böser, \textit{et al.}, \textbf{Prog in Particle and Nuclear Phy 111 (2020) 103736,} arXiv:1906.01739v3 [hep-ex].
	
	\bibitem{Ref20}	
	M. Cannoni, J. Ellis, M. Gómez and S. Lola, \textbf{Phys. Rev. D 88 (2013) 7, 075005,} [arXiv:hep-ph/1301.6002]. 
	
	\bibitem{Ref21}
	M. Gómez, G. Leontaris, S. Lola and J. Vergados, \textbf{Phys. Rev. D 59 (1999) 116009,} [arXiv:hep-ph/9810291].
	
	\bibitem{Ref22}
	J. Ellis, M. E. Gómez, G. Leontaris, S. Lola and D. Nanopoulos, \textbf{Eur. Phys. J. C 14 (2000) 319,} [arXiv:hep-ph/9911459].
	
	\bibitem{Ref23}
	J. Hisano, T. Moroi, K. Tobe and M. Yamaguchi, \textbf{Phys. Rev. D 53 (1996) 2442.}
	
	\bibitem{Ref24} Debtosh Chowdhury, Raghuveer Garani, Sudhir K. Vempati, \textit{ \texttt{SuSeFLAV 1.2}: Program for supersymmetric mass spectra with Seesaw Mechanism and rare lepton flavor violating decays,} \textbf{arXiv:1109.3551v5 [hep-ph]}.
	
	\bibitem{Ref25} H. P. Nilles, \textbf{Phys. Rept. 110 (1984) 1}.
	
	\bibitem{Ref26} G. D’Ambrosio \textit{et al.,} \textbf{Nucl. Phys. B 645 (2002) 155.} 
	
	\bibitem{Ref27} C. Smith, \textbf{Acta Phys. Polon. Supp. 3 (2010) 53}.
	
	\bibitem{Ref28} J. Ellis \textit{et. al}, \textbf{arXiv:2002.11057v1 [hep-ph]}.
	
	\bibitem{Ref29} B. Aubert, \textit{et al.,} \textbf{Phys. Rev. Lett. 104 (2010) 021802}, [arXiv:0908.2381].
	
	\bibitem{Ref30} K. Hayasaka, \textbf{Nucl. Phys. Proc. Suppl. 225-227 (2012) 169-172}. 
	
	\bibitem{Ref31} S. Dimopoulos and H. Georgi, \textbf{Nucl. Phys. B 193 (1981) 150.}
	
	\bibitem{Ref32} W. de Boer, \textbf{Prog. Part. Nucl. Phys. 33 (1994) 201-302,} [arXiv:hep-ph/9402266v5]. 
	
	\bibitem{Ref33} DELPHI Collaboration. \textbf{Eur.Phys.J.C 31 (2003) 421-479.}
	
	\bibitem{Ref34} G.L. Kane \textit{et al.}, \textbf{Phys. Rev.D 49 (1994) 6173.}
	
	\bibitem{Ref35} GAMBIT Collaboration. P. Athron, \textit{et al.} \textbf{Eur. Phys. J. C 77 (2017) 824.} 
	
	\bibitem{Ref36} CMS Collaboration. \textbf{Phys. Rev.D 90 (2014) 9 092007,} arXiv:1409.3168 [hep-ex].
	
	\bibitem{Ref37} K. T. Matchev and S. D. Thomas. \textbf{Phys. Rev. D 62 (2000) 077702,} arXiv:hep-ph/9908482.
	
	\bibitem{Ref38} CMS Collaboration. \textbf{Phys. Rev.D 100 (2019) 11, 112003,} arXiv:1909.06166 [hep-ex].
	
	\bibitem{Ref39} CMS Collaboration. \textbf{JHEP 03 (2018) 160,} arXiv: 1801.03957 [hep-ex].
	
	\bibitem{Ref40} CMS Collaboration. \textbf{JHEP 01 (2019) 154,} arXiv:1812.04066 [hep-ex]. 
	
\end{thebibliography}
\end{document}